\documentclass[a4paper,10pt]{article}
\usepackage{graphicx}
\usepackage{comment}
\usepackage{amssymb}
\usepackage{amsmath}
\usepackage{nicefrac}
\usepackage{graphicx}
\usepackage{dcolumn}
\usepackage{bm}
\usepackage{comment}
\usepackage{multirow}
\usepackage{cite}

\begin{document}

\huge

\begin{center}
Recursive determination of phase shifts for screened Coulomb potentials
\end{center}

\vspace{0.5cm}

\large

\begin{center}
Jean-Christophe Pain\footnote{jean-christophe.pain@cea.fr}
\end{center}

\normalsize

\begin{center}
\it CEA, DAM, DIF, F-91297 Arpajon, France
\end{center}

\vspace{0.5cm}

\begin{abstract}
In the calculation of hot-plasma atomic structure, the continuum wavefunctions are characterized by phase shifts, which therefore determine the scattering cross-sections. In this short paper, we propose a recurrence relation for the phase shifts in the case of a particular type of parametric potentials widely used in atomic-structure codes. These potentials have to be linear combinations of static screened Coulomb potentials (Yukawa-type potentials) multiplied by polynomial functions. 
\end{abstract}

\section{Introduction}

The potential felt by an electron in the atom is not coulombic at intermediate distances, and the difference between the effective potential and the Coulomb potential manifests itself through the quantum defect associated with bound states and the phase shift associated with continuum states. For a given angular momentum, the quantum defect $\gamma_{n\ell}$ \cite{SEATON83} is defined, in atomic units, by

\begin{equation}
\epsilon_{n\ell}=-\frac{Z^2}{2\left(n-\gamma_{n\ell}\right)^2},
\end{equation}

\noindent where $\epsilon_{n\ell}$ is the binding energy of the excited electron in the $n\ell$ orbital and $Z$ the charge of the ionic core. When the electron energy is above the continuum limit, its wavefunction is characterized by a phase shift $\delta_{\ell}$ for each orbital quantum number $\ell$. The asymptotic solution of the radial Schr\"odinger equation

\begin{equation}
\frac{d^2\psi_{\ell}}{dr^2}+\left(\epsilon-V(r)-\frac{\ell(\ell+1)}{r^2}\right)\psi_{\ell}=0
\end{equation}

\noindent has the form

\begin{equation}
\psi_{\ell}(kr)=C_{\ell}\sqrt{\frac{\pi kr}{2}}\left(\cos[\delta_{\ell}(k)]J_{\ell+1/2}(kr)-\sin[\delta_{\ell}(k)]J_{\ell-1/2}(kr)]\right),
\end{equation}

\noindent where $C_{\ell}$ is a constant, $k=\sqrt{2\epsilon}$ and $J_{\ell+1/2}$ and $J_{\ell-1/2}$ are Bessel functions of the first kind. When $r\rightarrow \infty$, $\psi_{\ell}$ has the asymptotic form

\begin{equation}\label{asymp}
\psi_{\ell}(kr)\approx C_{\ell}\sin\left[kr-\frac{\ell\pi}{2}+\delta_{\ell}(k)\right].
\end{equation}

\noindent The difference between the actual and the coulombic phase shift is the signature of the non-coulombic part of the electron-ion interaction. Phase shifts are important for the photo-ionization cross-sections and a reliable calculation of the impact ionization cross-section of an atom requires an accurate determination of the continuum wavefunctions in the incident and in the exit channels. In the theory of collisions, phase shifts determine the scattering cross-sections \cite{MURILLO98,SMIRNOV03}. For instance, the elastic scattering cross-section of particles is 

\begin{equation}
\sigma_s=\frac{4\pi}{k^2}\sum_{\ell}\left(2\ell+1\right)\sin^2\left[\delta_{\ell}(k)\right].
\end{equation}

\noindent The phase shift plays a major role in phase-amplitude methods \cite{BABIKOV67,NIKIFOROV05,BARSHALOM96,PAIN99,TANNOUS08}; but whatever the technique chosen for its computation (solving Schr\"odinger equation or using semi-classical methods \cite{WEI06}), the difficult point is the determination of the potential $V(r)$. Due to configuration mixing, the number of involved radial integrals can be very large, and analytical potentials can be an alternative to self-consistent field methods. Tannous \emph{et al.} \cite{TANNOUS99} tried to determine potentials which incorporate the effect of exchange while keeping a local character. Parametric potentials are often used \cite{MARTEL95}, and are the key ingredient of a number of atomic-structure codes, such as HULLAC (Hebrew University Lawrence Livermore Atomic Code) \cite{BARSHALOM01,BUSQUET06,KLAPISCH07,KLAPISCH09}, OPAL \cite{ROGERS88}, STA (Super Transition Arrays) \cite{BAR89,BAR06,BAR07} or FAC (Flexible Atomic Code) \cite{GU08}. In this work, we propose, following Tietz \cite{TIETZ63,SHELDON64}, to evaluate the difference between phase shifts associated to consecutive values of $\ell$ by the relation

\begin{equation}\label{eqa}
\sin\left[\delta_{\ell}(k)-\delta_{\ell+1}(k)\right]=\frac{\pi}{2k}\int_0^{\infty}r\frac{dV}{dr}J_{\ell+1/2}(kr)J_{\ell+3/2}(kr)dr,
\end{equation}

\noindent which is obtained by replacing the wavefunction by its asymptotic expression (\ref{asymp}). Assuming in addition that the difference $\delta_{\ell}(k)-\delta_{\ell+1}(k)$ is small yields 

\begin{equation}\label{eqab}
\delta_{\ell}(k)-\delta_{\ell+1}(k)=\frac{\pi}{2k}\int_0^{\infty}r\frac{dV}{dr}J_{\ell+1/2}(kr)J_{\ell+3/2}(kr)dr.
\end{equation}

Two parametric potentials used in widely-used atomic-structure codes are presented in Secs. \ref{klapi} and \ref{rog}, and the new recursion relation is explained in Sec. \ref{recu}.

\section{Klapisch's parametric potentials}\label{klapi}

The electronic shell structure in an atomic potential was introduced by Klapisch \cite{KLAPISCH69}. The parametric potential method consists in optimizing an analytical central-field potential according to a chosen quality criterion (variational, spectroscopic, \emph{etc}.). The method is particularly interesting, compared to Hartree-Fock for example, when the wavefunctions are nearly solutions of a central field (this is the case for alkali-like spectra or for highly ionized atoms) or if the spin-orbit is strong (rare gases, medium and heavy atoms) \cite{FENEUILLE36,AYMAR70}. It was implemented in the RELAC code \cite{LUC72,KLAPISCH77} and was widely used, for instance in the calculation of photo-ionization cross-sections \cite{AYMAR80,BARSHALOM83,AYMAR84,AYMAR96}. Let us consider a closed electronic subshell of $q$ electrons, having spherical symmetry, described by the normalized radial density of charge:

\begin{equation}
-q\frac{\alpha^{2\ell+3}}{(2\ell+2)!}\left(r^{\ell+1}e^{-\alpha r/2}\right)^2,
\end{equation}

\noindent where $\alpha$ is a positive real constant ($\alpha$ plays the role of the inverse of a screening length). The potential of another electron at radius $r$ in the field of the latter density of charge and a nucleus of charge $Z$ reads \cite{KLAPISCH71}

\begin{equation}
V(\alpha,r)=-\frac{1}{r}\left[qf(\ell,\alpha,r)+Z-q\right],
\end{equation}

\noindent where

\begin{equation}\label{klap}
f(\ell,\alpha,r)=e^{-\alpha r}\sum_{j=0}^{2\ell+1}\left(1-\frac{j}{2\ell+2}\right)\frac{(\alpha r)^j}{j!}.
\end{equation}

\noindent To each subshell $i$ corresponds a parameter $\alpha_i$. The parameters $\alpha_i$, $i\in[1,N_s]$, $N_s$ being the number of subshells, are adjusted by a least-square procedure to reproduce experimental data or \emph{ab initio} calculations. The function $f(\ell,\alpha,r)$ of Eq. (\ref{klap}) can be expressed in terms of incomplete Gamma functions (see \ref{appa}), which can be expressed in terms of confluent hypergeometric functions (or ``Kummer functions'') $~_1F_1$. The first term in the summation ($j$=0) corresponds to the Yukawa potential \cite{YUKAWA35}, for which it is difficult to obtain an analytical treatment of bound states, although significant progress was made in the last few years (see for instance Ref. \cite{BAHLOULI10}). 

Historically, Yukawa showed in the 1930s that such a potential arises from the exchange of a massive scalar field such as 
the field of a massive boson. Since the field mediator is massive, the corresponding force has a certain range, which 
is inversely proportional to the mass of the mediator particle \cite{MULLER68}. 

The screening (or shielding) effect describes the attraction between an electron and the nucleus in any atom with more than one electron. It can be defined as a reduction in the effective nuclear charge felt by an electron, due to its interactions with the other electrons and to the interactions of these electrons with the nucleus. In low-density, high-temperature plasmas, electric-field screening can be taken into account using the Debye-H\"uckel model \cite{DEBYE23}. 

In solid-state physics, the screened potential is used to calculate the electronic band structure of a large variety of materials, often in combination with pseudopotential models.

\section{Parametric potential of Rogers \emph{et al.}}\label{rog}

Rogers \emph{et al.} \cite{ROGERS88,ROGERS81} defined the electron configuration as having two components: the first one is a parent configuration consisting of all the electrons in a given configuration except one. The excluded electron defines the second component or ``running'' electron. The authors introduced a parametric potential for each parent configuration. This potential consists of a long-range Coulomb part and a screened function represented by a sum of static screened Coulomb (Yukawa-type) potentials \cite{YUKAWA35}:

\begin{equation}
V(r)=-\frac{1}{r}\left[(Z-\nu)+\sum_{n=1}^{n^*}N_n~e^{-\alpha_n r}\right],
\end{equation}

\noindent where

\begin{equation}
\nu=\sum_{n=1}^{n^*}N_n
\end{equation}

\noindent is the number of electrons in the parent ion, $N_n$ the number of electrons in the shell with principal quantum number $n$, $n^*$ the maximum value of $n$ for the parent configuration and $\alpha_n$ the screening constant for electrons in shell $n$. For a given ion, it is possible to define as many parent configurations as occupied subshells. The screening parameters are obtained by an iterative procedure consisting in the resolution of spin-averaged Dirac equation and matching the eigenvalues to the ionization potentials. Such an approach enables one to account for effects (electron correlations, spin-other-orbit interactions) which are not necessarily included in \emph{ab initio} calculations. The authors provided accurate fits of the screening parameters along an iso-electronic sequence. The estimation of the screening constants was improved by Mabong \emph{et al.} \cite{MABONG96a} in order to include relativistic effects and later by Mendoza \emph{et al.} \cite{MENDOZA11} and others. Rogers' potential encountered a great success for the generation of atomic quantities (levels, oscillator strengths, photo-ionization cross-sections, etc.) required in astrophysics. The potential can be applied to any excitation and ionization state. Moreover, since its Fourier transform is simple, it is a good candidate for the determination of cross-sections of elementary processes.

In the present work we only consider the static screening effects. It is however important to keep in mind that the dynamic screening effects are also important not only in weakly coupled plasmas but also in strongly coupled plasmas such as a dusty plasmas \cite{VRANJES02,RAMAZANOV08,KI12}. Many interesting studies were (and are being) carried out on the subject (time-dependent density functional theory, linear response theory, etc.), but including the effect of dynamic screening properly in our model would be a difficult task and is beyond the scope of the present article. In the framework of the linear response theory, Zangwill and Soven \cite{ZANGWILL80} applied the random-phase-approximation to calculate the photo-absorption cross-section of rare gases. This was the starting point for the development of the time-dependent density functional theory. Extending the scope of this work to plasmas leads to some difficulties because the atoms cannot be considered without delocalized electrons. The latter may contribute to particle-hole transitions but may also be responsible for collective effects. A proper description of dynamic screening would then require a full quantum-mechanical description of both bound and free electronic states. The effect of dynamic screening on the energies of the electronic subshells would be difficult to quantify but we expect it to be rather small. Dynamical screening is known to have an impact on the radiative spectra [see for instance \cite{ANKUDINOV03}, through channel mixing (photo-excitation / photo-ionization) and configuration interaction.


\section{Recursion relation for the phase shift}\label{recu}

The recursion evaluation of the phase shift following the requirements of Tietz \cite{TIETZ63} (see Eq. (\ref{eqa})) involves only integrals of the type

\begin{equation}\label{bas}
\int_0^{\infty}x^{\lambda-1}e^{-\alpha x}J_{\mu}(\beta x)J_{\nu}(\gamma x)dx,
\end{equation}

\noindent where $J_{\mu}$ and $J_{\nu}$ are Bessel functions of the first kind. Although efficient algorithms do exist (for instance Lucas \cite{LUCAS95} proposed a method which makes use of extrapolation on a sequence of partial sums, and requires rewriting the product of Bessel functions as the sum of two more well-behaved functions), the evaluation of infinite integrals involving products of Bessel functions is tedious. The integral (\ref{bas}) can be expressed, if $\lambda+\mu+\nu>0$ and $\alpha>0$, as (see Eq. (6.626) of Ref. \cite{GRADSHTEYN80} p. 715):

\begin{eqnarray}
\int_0^{\infty}x^{\lambda-1}e^{-\alpha x}J_{\mu}(\beta x)J_{\nu}(\gamma x)dx&=&\frac{\beta^{\mu}\gamma^{\nu}}{2^{\mu+\nu}\alpha^{\mu+\nu+\lambda}\Gamma(\nu+1)}\nonumber\\
& &\times\sum_{m=0}^{\infty}\frac{\Gamma(\lambda+\mu+\nu+2m)}{m!~\Gamma(\mu+m+1)}\left(-\frac{\beta^2}{4\alpha^2}\right)^m\nonumber\\
& &\times~_2F_1\left(-m,-\mu-m;\nu+1;\frac{\gamma^2}{\beta^2}\right),\nonumber\\
& &
\end{eqnarray}

\noindent where $\Gamma$ is the usual Gamma function and $_pF_q$ is a generalized hypergeometric function defined by

\begin{eqnarray}\label{def_fpq}
_pF_q\left[\begin{array}{c}
\alpha_1\cdots\alpha_p\\
\gamma_1,\cdots,\gamma_q
\end{array};z\right]&=&\sum_{n=0}^{\infty}\frac{\prod_{i=1}^p(\alpha_i)_n}{\prod_{j=1}^q(\gamma_j)_n}\frac{z^n}{n!},
\end{eqnarray}

\noindent where $(\lambda)_n$ represents the Pochhammer symbol

\begin{equation}
(\lambda)_n=\lambda(\lambda+1)\cdots(\lambda+n-1)=\frac{\Gamma(\lambda+n)}{\Gamma(\lambda)}.
\end{equation}

\noindent In particular, $_1F_1$ is the confluent hypergeometric function, which includes Bessel functions, Laguerre polynomials, incomplete Gamma functions, \emph{etc.} as special cases. The function $_2F_1$, commonly referred to as the Gauss hypergeometric series, includes Chebyshev, Legendre and Jacobi polynomials. In the particular case where $\beta=\gamma=k$, we find, after some algebraic manipulation, the following identity:

\begin{eqnarray}
& &\int_0^{\infty}x^{\lambda-1}e^{-\alpha x}J_{\mu}(kx)J_{\nu}(kx)dx\nonumber\\
& &\;\;\;\;\;\;\;\;\;\;\;\;\;\;=\frac{k^{\mu+\nu}}{2^{\mu+\nu}\alpha^{\lambda+\mu+\nu}}\frac{\Gamma(\mu+\nu+\lambda)}{\Gamma(\mu+1)\Gamma(\nu+1)}\nonumber\\
& &\;\;\;\;\;\;\;\;\;\;\;\;\;\;\times~_4F_3\left[\begin{array}{c}
\frac{\mu+\nu+1}{2},\frac{\mu+\nu+2}{2},\frac{\mu+\nu+\lambda}{2},\frac{\mu+\nu+\lambda+1}{2}\\
\mu+1,\nu+1,\mu+\nu+1
\end{array};-\frac{4k^2}{\alpha^2}\right].
\end{eqnarray}

\noindent In our case, since we have $\mu=\ell+1/2$ and $\nu=\ell+3/2$, the $_4F_3$ generalized hypergeometric function reduces to $_3F_2$ (namely ``Clausen function''), which leads to the final expression

\begin{eqnarray}\label{fid}
& &\int_0^{\infty}x^{\lambda-1}e^{-\alpha x}J_{\ell+1/2}(kx)J_{\ell+3/2}(kx)dx\nonumber\\
& &\;\;\;\;\;\;\;\;\;\;\;\;\;\;=\frac{k^{2\ell+2}}{\sqrt{\pi}\alpha^{2\ell+2+\lambda}}\frac{\Gamma(\ell+2)\Gamma(2\ell+2+\lambda)}{\Gamma(\ell+\frac{5}{2})\Gamma(2\ell+3)}\nonumber\\
& &\;\;\;\;\;\;\;\;\;\;\;\;\;\;\times~_3F_2\left[\begin{array}{c}
\ell+2,\ell+1+\frac{\lambda}{2},\ell+\frac{3}{2}+\frac{\lambda}{2}\\
\ell+\frac{5}{2}, 2\ell+3
\end{array};-\frac{4k^2}{\alpha^2}\right]\nonumber\\
& &\;\;\;\;\;\;\;\;\;\;\;\;\;\;\equiv I_{\lambda}(k,\alpha,\ell).
\end{eqnarray}

\noindent This result was also obtained by Mahajan \cite{MAHAJAN75}. A similar integral, involved in the difference between $\delta_{\ell-1}(k)$ and $\delta_{\ell+1}(k)$ is given in \ref{appb}. When $\alpha=0$ and $\lambda=0$, one has

\begin{equation}
I_{0}(k,0,\ell)=\int_0^{\infty}\frac{1}{x}J_{\ell+1/2}(kx)J_{\ell+3/2}(kx)dx=\frac{1}{\pi(\ell+1)}.
\end{equation}

\noindent The numerical evaluation of hypergeometric functions $_pF_q$ was investigated by many authors. For instance, the program of Perger \emph{et al.} \cite{PERGER93} relies on a direct numerical evaluation of the series. The only transformation formula employed is the use of the linear transformation

\begin{eqnarray}
_2F_1(a,b;c;z)&=&\frac{\Gamma(c)\Gamma(c-a-b)}{\Gamma(c-a)\Gamma(c-b)}~_2F_1(a,b;a+b-c+1;1-z)\nonumber\\
& &+(1-z)^{c-a-b}\frac{\Gamma(c)\Gamma(a+b-c)}{\Gamma(a)\Gamma(b)}\nonumber\\
& &\times~_2F_1(c-a,c-b;c-a-b+1;1-z)
\end{eqnarray}

\noindent when $|z|$ is approaching 1 from below. Recently, Willis \cite{WILLIS12} proposed an acceleration procedure through precise remainder asymptotics. He expressed the asymptotics of the remainders of the partial sums of the generalized hypergeometric function $_pF_q$ through an inverse power series

\begin{equation}
z^nn^{\lambda}\sum_k\frac{c_k}{n^k},
\end{equation}

\noindent where the exponent $\lambda$ and the asymptotic coefficients $c_k$ may be recursively computed to any desired order from the hypergeometric parameters and argument. The hypergeometric series $_3F_2$ can also be evaluated numerically using the following integral representation \cite{KAMPE26,KAMPE37}:
 
\begin{equation}
~_3F_2\left[\begin{array}{c}
a,b,c\\
d, e
\end{array};x\right]=C\int_0^1\int_0^1y^{a-1}(1-y)^{d-a-1}z^{b-1}(1-z)^{e-b-1}(1-yz)^{-c}dydz, 
\end{equation}

\noindent where

\begin{equation}
C=\frac{\Gamma(d)\Gamma(e)}{\Gamma(a)\Gamma(d-a)\Gamma(b)\Gamma(e-b)}
\end{equation}

\noindent with $d-a\ge 0$ and $e-b\ge 0$. Such a representation is interesting since the integral over the whole space (from 0 to $\infty$) of the left-hand side of Eq. (\ref{fid}) is replaced by a finite-range integral between 0 and 1. Other numerical techniques exist. For example, Wills \cite{WILLS71} and a few years later Bretz \cite{BRETZ76}, arranged the series expansion into a nested form. The ease of computing numerically the $_pF_q$ using Horner's rule \cite{LEE66} for polynomial evaluation has been found to yield fortran programs which are very accurate (see for instance the program by Srinivasa Rao in Ref. \cite{RAO78}). The generalized hypergeometric function is evaluated as

\begin{equation}
~_pF_q(\alpha_1,\alpha_2,\cdots,\alpha_p;\beta_1,\beta_2,\cdots,\beta_q;Z)=\left[1+\frac{X_0}{Y_0}\left(Z+\frac{X_1}{Y_1}\left(Z+\frac{X_2}{Y_2}+\cdots\right)\right)\right],
\end{equation}

\noindent with

\begin{equation}
X_i=\prod_{j=1}^p\left(\alpha_j+i\right)
\end{equation}

\noindent and

\begin{equation}
Y_i=(i+1)\prod_{k=1}^q\left(\beta_k+i\right).
\end{equation}

\noindent However, we do not pretend that the numerical evaluation of hypergeometric functions is more efficient from a numerical point of view (speed, accuracy) than direct integration. Such considerations depend on the numerical integration technique used, on the machine, etc. The purpose of the present work is to provide analytical expressions, which interest, beyond the purely mathematical aspect, relies in the fact that they can open the way to algebraic manipulations, through the use of properties, recurrence relations, etc. of hypergeometric functions. 

Let us consider the static screened Coulomb potential

\begin{equation}\label{stscr}
V(r)=-\frac{e^{-\alpha r}}{r},
\end{equation}

\noindent for which the calculation of the phase shifts was investigated by many authors. Rogers \cite{ROGERS71} proposed a direct integration near the origin and first- and second-order WKB approximation ar larger distance. Grandjouan and Deutsch \cite{GRANDJOUAN75} used Numerov method and variable-phase approach, applied to the Sturm-Liouville form of the Schr\"odinger equation. Bechler and Pratt \cite{BECHLER87} performed the calculation in the framework of perturbation theory. Following the prescription of Tietz \cite{TIETZ63}, we get

\begin{equation}\label{yuk}
\delta_{\ell}(k)-\delta_{\ell+1}(k)=\frac{\pi}{2k}\left[\alpha I_1(k,\alpha,\ell)+I_0(k,\alpha,\ell)\right]
\end{equation}

\noindent or equivalently

\begin{equation}\label{yuk2}
\delta_{\ell}(k)=\delta_0(k)-\frac{\pi}{2k}\sum_{p=0}^{\ell-1}\left[\alpha I_1(k,\alpha,p)+I_0(k,\alpha,p)\right].
\end{equation}

\clearpage

\begin{table}[!ht]
\caption{\label{tab1} Numerical values for the potential parameters $\gamma_i$ and $\lambda_i$ for argon (from Ref. \cite{TIETZ65} for the potential given by Eq. (\ref{t65})).}
\begin{tabular}{lllllll}\hline\hline
 $\gamma_1$ & $\gamma_2$ & $\gamma_3$ & $\lambda_1$ & $\lambda_2$ & $\lambda_3$ \\\hline
 0.50529 & 0.43447 & 0.06071 & 2.68764 & 9.06392 & 46.49853 \\\hline\hline
\end{tabular}
\end{table}

\begin{table}[!ht]
\caption{\label{tab2} Consecutive phase-shift differences $\delta_{\ell}-\delta_{\ell+1}$ for argon and $E$=$k^2/2$=40 keV with the potential given by Eq. (\ref{t65}) with parameters of table \ref{tab1}. Comparison between the present work and the numerical integration.}
\begin{tabular}{llll}\hline\hline
 $\ell$ & Num. int. & This work & Rel. diff. (\%)\\\hline
 0 & 0.3412 & 0.31945 & 6.4\\
 1 & 0.1732 & 0.1512 & 12.7\\
 2 & 0.1110 & 0.0953 & 14.1\\ 
 3 & 0.0824 & 0.0675 & 18.1\\ 
 4 & 0.0630 & 0.0510 & 19.0\\ 
 5 & 0.0510 & 0.0402 & 21.2\\\hline\hline
\end{tabular}
\end{table}

\begin{table}[!ht]
\caption{\label{tab3} Numerical values for the potential parameters $\gamma_i$ and $\lambda_i$ for mercury (from Ref. \cite{ROZENTAL36,TIETZ59} with the potential given by Eq. (\ref{t65})).}
\begin{tabular}{lllllll}\hline\hline
 $\gamma_1$ & $\gamma_2$ & $\gamma_3$ & $\lambda_1$ & $\lambda_2$ & $\lambda_3$ \\\hline
 0.255 & 0.581 & 0.164 & 0.246 & 0.947 & 4.356 \\\hline\hline
\end{tabular}
\end{table}

\begin{table}[!ht]
\caption{\label{tab4} Consecutive phase-shift differences $\delta_{\ell}-\delta_{\ell+1}$ for mercury ($Z$=80) with $k=Z$ with the potential given in Ref. \cite{ROZENTAL36,TIETZ59} (see table \ref{tab3}), which was proposed a long time ago as a parametrization of the Thomas-Fermi potential. Comparison between the present approach and the numerical integration.}
\begin{tabular}{llll}\hline\hline
 $\ell$ & Num. int. & This work & Rel. diff. (\%)\\\hline
 1 & 0.4742 & 0.4982 & 5.1\\
 2 & 0.3200 & 0.3310 & 3.4\\
 3 & 0.2356 & 0.2473 & 5.0\\ 
 4 & 0.1774 & 0.1969 & 10.9\\ 
 5 & 0.1426 & 0.1632 & 14.4\\\hline\hline
\end{tabular}
\end{table}

\begin{figure}
\begin{center}
\vspace{1cm}
\includegraphics[width=8cm]{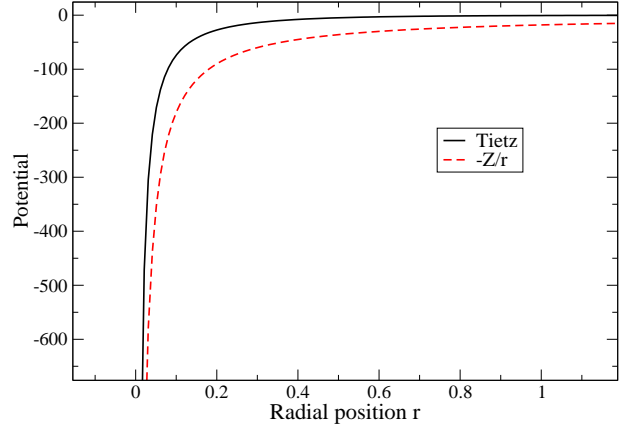}
\end{center}
\caption{Potential used in Ref. \cite{TIETZ65} (the parameters are provided in table \ref{tab1}).}\label{fig1}
\end{figure}

\vspace{1cm}

\begin{figure}
\begin{center}
\vspace{0.1cm}
\includegraphics[width=8cm]{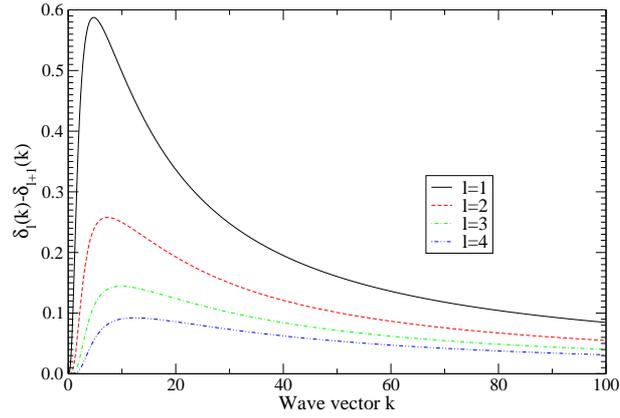}
\end{center}
\caption{Variation of $\delta_{\ell}(k)-\delta_{\ell+1}(k)$ with respect to $k$ using Eq. (\ref{delti}) and the potential used in Ref. \cite{TIETZ65} (the parameters are provided in table \ref{tab1}).}\label{fig2}
\end{figure}

\clearpage

\begin{table}[!ht]
\caption{\label{tab5} Numerical values for the potential parameters $\gamma_i$ and $\lambda_i$ for uranium (from Ref. \cite{KARLE64} for the potential given by Eq. (\ref{t65})).}
\begin{tabular}{lllllll}\hline\hline
 $\gamma_1$ & $\gamma_2$ & $\gamma_3$ & $\lambda_1$ & $\lambda_2$ & $\lambda_3$ \\\hline
 0.31000 & 0.56667 & 0.12346 & 2.9802 & 10.564 & 50.463 \\\hline\hline
\end{tabular}
\end{table}

\begin{table}[!ht]
\caption{\label{tab6} Consecutive phase-shift differences $\delta_{\ell}-\delta_{\ell+1}$ for uranium and $E$=$k^2/2$=40 keV with the potential given by Eq. (\ref{t65}) and the parameters of table \ref{tab5}. Comparison between the present work and the numerical integration (denoted ``Num. int.'').}
\begin{tabular}{llll}\hline\hline
 $\ell$ & Num. int. & This work & Rel. diff. (\%) \\\hline
 1 & 0.7856 & 0.7731 & 1.6\\
 2 & 0.5530 & 0.4871 & 11.9\\ 
 3 & 0.4085 & 0.3452 & 15.5\\ 
 4 & 0.3110 & 0.2608 & 16.1\\ 
 5 & 0.2480 & 0.2053 & 17.2\\ 
 6 & 0.1989 & 0.1661 & 16.5\\ 
 7 & 0.1690 & 0.1372 & 18.8\\ 
 8 & 0.1400 & 0.1152 & 17.7\\ 
 9 & 0.1209 & 0.0980 & 18.9\\ 
10 & 0.1029 & 0.0843 & 18.1\\ 
11 & 0.0920 & 0.0731 & 20.5\\ 
12 & 0.0790 & 0.0640 & 19.0\\ 
13 & 0.0710 & 0.0563 & 20.7\\ 
14 & 0.0610 & 0.0499 & 18.2\\ 
15 & 0.0560 & 0.0445 & 20.5\\ 
16 & 0.0510 & 0.0398 & 22.0\\ 
17 & 0.0450 & 0.0358 & 20.4\\ 
18 & 0.0410 & 0.0323 & 21.2\\ 
19 & 0.0371 & 0.0293 & 21.0\\\hline\hline
\end{tabular}
\end{table}

\noindent In order to test the validity of the approximation, let us take the following potential (see Fig. \ref{fig1}) from Ref. \cite{TIETZ65}:

\begin{equation}\label{t65}
V(r)=-\frac{2Z}{r}\left[\sum_{i=1}^3\gamma_i\exp\left( -\lambda_ir\right)\right]
\end{equation}

\noindent with the parameters given in table \ref{tab1} for the case of argon, in table \ref{tab3} for mercury and in table \ref{tab5} for uranium. The integral (\ref{fid}) must be evaluated for the three different pairs $\left(\gamma_i,\lambda_i\right)$:

\begin{equation}\label{delti}
\delta_{\ell}(k)-\delta_{\ell+1}(k)=\frac{\pi Z}{k}\sum_{i=1}^3\gamma_i\left[\lambda_iI_1(k,\lambda_i,\ell)+I_0(k,\lambda_i,\ell)\right].
\end{equation}

\noindent We can see in tables \ref{tab2}, \ref{tab4} and \ref{tab6} that the values of the phase shift differences are rather close to the values obtained by numerical integration of Schr\"odinger equation. The third column contains the relative difference (denoted ``Rel. diff.'') defined as

\begin{equation}
\mathrm{Rel.~diff.}=\frac{\left|\mathrm{Num.~int.} - \mathrm{This~work}\right|}{\mathrm{Num.~int.}}.
\end{equation}

\noindent The numerical integration technique is described in Ref. \cite{KARLE64}; it consists in a modification of the Gauss-Jackson method \cite{MOTT49}. For mercury and uranium, the relativistic effects can not be neglected. Therefore, due to the high values of the atomic numbers (80 and 92 respectively), a proper description of the atomic structure would require solving Dirac's equation. However, in that case, a generalization of the present approach would be more complicated (in particular because the wavefunction has two components). When relativity is taken into account in the calculation of atomic structure, the energies of $s$ and $p$ orbitals become lower than in the non-relativistic case, on the contrary to $d$ and $f$ orbitals. The spin-orbit splitting is a very important relativistic effect. It concerns only non-$s$ orbitals. For $s$ orbitals the mass-velocity correction dominates over the Darwin correction, while for non-$s$ orbitals the Darwin correction is small. For inner-core $s$ and $p$ orbitals, the contraction of the orbital is easily explained by the relativistically increased mass. For instance, the energy of $1s$ orbital of uranium is (in atomic units): -3690.78 in the non-relativistic (Schr\"odinger) case, -4255.56 in the relativistic (Dirac) case, and -4114.71 in the semi-relativistic Pauli approximation. However, in the present work, we are dealing with continuum (free-electron) wavefunctions which are much less sensitive to relativistic effects than the bound states. Moreover, the relativistic effects become less and less important as the energy increases, and all our examples correspond to very high values of continuum energy. We are presenting an approximate method, in which the wavefunctions are represented by Bessel functions and we are only interested in phase shifts. The potential error resulting from the fact that we neglect relativistic effects is much smaller than the one resulting from the approximations mentioned above. 

Using the function ``Timing'' of the software Mathematica \cite{MATHEMATICA}, we find that our approach (see formula (\ref{delti})) is much more efficient than the direct numerical integration of the expression given in Eq. (\ref{eqab}). Different time evaluations are given in table \ref{tab7} for different values of atomic number $Z$ and orbital quantum number $\ell$. As expected, the numerical cost of the direct integration increases with $\ell$.

\begin{table}[!ht]
\caption{\label{tab7} Calculation time of onsecutive phase-shift differences $\delta_{\ell}-\delta_{\ell+1}$ for several values of $Z$ and $\ell$: comparison between the present approach (Eq. (\ref{delti})) and the numerical integration (Eq. (\ref{eqab})).}
\begin{tabular}{llll}\hline\hline
 Atomic  & Orbital & Time (s)              & Time (s)\\
 number  & number  & Formula (\ref{delti}) & Numerical\\
 $Z$     & $\ell$  &                       & integration (Eq. (\ref{eqab}))\\\hline
 18 & 1 & 0.30 & 1.95\\
 18 & 5 & 0.29 & 4.21\\
 18 & 10 & 0.29 & 7.19\\
 80 & 1 & 0.28 & 1.97\\
 80 & 5 & 0.28 & 4.43\\
 80 & 10 & 0.29 & 6.96\\ 
 92 & 1 & 0.27 & 2.00\\ 
 92 & 5 & 0.28 & 4.16\\
 92 & 10 & 0.29 & 7.39\\\hline\hline
\end{tabular}
\end{table}

As explained above, Tietz approximation relies on two assumptions: the wavefunctions are replaced by their approximate form and the phase shift differences must be small. The later approximation is not necessary; indeed, instead of Eq. (\ref{eqa}) we can take 

\begin{equation}
\delta_{\ell}(k)-\delta_{\ell+1}(k)=\arcsin\left[\frac{\pi}{2k}\int_0^{\infty}r\frac{dV}{dr}J_{\ell+1/2}(kr)J_{\ell+3/2}(kr)dr\right],
\end{equation}

\noindent which can be evaluated in the samme manner. The phase-shift difference for $\ell$=2, 3 and 4 is represented in figure \ref{fig2}. In the case of Klapisch's potential, the integral (\ref{fid}) must be evaluated for different values of $\lambda$, from 0 to $2\ell+2$ for a $n\ell$ subshell:

\begin{eqnarray}
\delta_{\ell}(k)-\delta_{\ell+1}(k)&=&\frac{\pi}{2k}(Z-q)I_0(k,0,\ell)+\frac{\pi}{2k}q\sum_{j=0}^{2\ell+1}\left(1-\frac{j}{2\ell+2}\right)\nonumber\\ 
& &\times\frac{\alpha^j}{j!}\left[I_{j+1}(k,\alpha,\ell)\alpha-(j-1)I_j(k,\alpha,\ell)\right].\nonumber\\
& &
\end{eqnarray}

\noindent As concerns, Rogers' potential, since it is built as a linear combination of Yukawa potentials, the result is directly given by Eq. (\ref{yuk}) evaluated $n^*$ times for different sets of parameters:

\begin{eqnarray}
\delta_{\ell}(k)-\delta_{\ell+1}(k)&=&\frac{\pi}{2k}(Z-\nu)I_0(k,0,\ell)\nonumber\\ 
& &+\frac{\pi}{2k}\sum_{n=1}^{n^*}N_n\left[I_0(k,\alpha_n,\ell)-\alpha_nI_1(k,\alpha_n,\ell)\right].
\end{eqnarray}

\section{Conclusion}

Starting from Tietz formula, in which the wavefunctions are replaced by their asymptotic form, we proposed an analytic expression for the difference between two consecutive phase shifts (in terms of orbital quantum number $\ell$) for a class of screened Coulomb potentials. This recursion relation, which involves a $_3F_2$ generalized hypergeometric function, enables one to obtain approximate phase shift in atomic structure codes using parametric potentials consisting of the summation of a pure Coulomb part and static screened Coulomb potentials multiplied by polynomial functions.

\section{Acknowledgments}

The author would like to thank D. Teychenn\'e and A. Decoster for useful discussions about the numerical evaluation of the generalized hypergeometric functions.


\appendix

\section{Expression of Klapisch's potential in terms of incomplete Gamma functions}\label{appa}

The potential reads

\begin{equation}
V(\alpha,r)=-\frac{1}{r}\left[qf(\ell,\alpha,r)+Z-q\right],
\end{equation}

\noindent with

\begin{eqnarray}\label{klap2}
f(\ell,\alpha,r)&=&e^{-\alpha r}\sum_{j=0}^{2\ell+1}\left(1-\frac{j}{2\ell+2}\right)\frac{(\alpha r)^j}{j!}\nonumber\\
&=&e^{-\alpha r}\sum_{j=0}^{2\ell+1}\frac{(\alpha r)^j}{j!}-\alpha r\frac{e^{-\alpha r}}{2\ell+2}\sum_{j=0}^{2\ell}\frac{(\alpha r)^j}{j!}.
\end{eqnarray}

\noindent Since we have

\begin{equation}
e^{-x}\sum_{j=0}^{2\ell+1}\frac{x^j}{j!}=\frac{\tilde{\Gamma}\left(2\ell+2,x\right)}{\tilde{\Gamma}\left(2\ell+2\right)},
\end{equation}

\noindent where $\tilde{\Gamma}$ is the incomplete Gamma function

\begin{equation}
\Gamma\left(a,x\right)=\int_x^{\infty}e^{-t}t^{a-1}dt,
\end{equation}

\noindent equation ({\ref{klap2}) becomes

\begin{equation}
f(\ell,\alpha,r)=\frac{1}{\tilde{\Gamma}\left(2\ell+1\right)}\left[\frac{(2\ell+2-\alpha r)}{(2\ell+2)}\tilde{\Gamma}\left(2\ell+1,\alpha r\right)+\frac{\left(\alpha r\right)^{2\ell+1}}{(2\ell+1)}e^{-\alpha r}\right].
\end{equation}

\noindent Using

\begin{equation}
\Gamma\left(a+1,x\right)=a\tilde{\Gamma}\left(a,x\right)+x^ae^{-x},
\end{equation}

\noindent we finally obtain

\begin{equation}
f(\ell,\alpha,r)=\frac{1}{\tilde{\Gamma}\left(2\ell+2\right)}\left[\frac{(2\ell+1)(2\ell+2-\alpha r)}{(2\ell+2)}\tilde{\Gamma}\left(2\ell+1,\alpha r\right)+\left(\alpha r\right)^{2\ell+1}e^{-\alpha r}\right].
\end{equation}

\section{Sheldon's formula for non-consecutive phase shifts}\label{appb}

We believe it is important to recall the reader

\begin{equation}
\delta_{\ell}(k)-\delta_{\ell+1}(k)=\frac{\pi}{2k}\int_0^{\infty}r\frac{dV}{dr}J_{\ell+1/2}(kr)J_{\ell+3/2}(kr)dr.
\end{equation}

\noindent Using the following relation for the Bessel functions

\begin{equation}
J_{n+1}(z)+J_{n-1}(z)=\frac{2n}{z}J_n(z),
\end{equation}

\noindent Tietz \cite{TIETZ63} obtained the following expression

\begin{equation}
\delta_{\ell-1}(k)-\delta_{\ell+1}(k)=\frac{\left(\ell+\frac{1}{2}\right)\pi}{k^2}\int_0^{\infty}\frac{dV}{dr}\left[J_{\ell+1/2}(kr)\right]^2dr.
\end{equation}

\noindent In order to obtain an expression depending not only on the derivative of the potential but on the potential itself, Sheldon suggested to perform an integration by part

\begin{eqnarray}
\delta_{\ell-1}(k)-\delta_{\ell+1}(k)&=&-\int_0^{\infty}\frac{2\pi\left(\ell+\frac{1}{2}\right)}{k}J_{\ell+1/2}(kr)J_{\ell-1/2}(kr)V(r)dr\nonumber\\
& &+\int_0^{\infty}\frac{2\pi\left(\ell+\frac{1}{2}\right)^2}{k^2r}\left[J_{\ell+1/2}(kr)\right]^2dr.
\end{eqnarray}

\noindent Using the following identity: 

\begin{eqnarray}
\left[J_{\ell+1/2}(kr)\right]^2-\left[J_{\ell-1/2}(kr)\right]^2&=&\frac{4\left(\ell+\frac{1}{2}\right)^2}{(kr)^2}\left[J_{\ell+1/2}(kr)\right]^2\nonumber\\
& &-\frac{4\left(\ell+\frac{1}{2}\right)}{kr}J_{\ell+1/2}(kr)J_{\ell-1/2}(kr),\nonumber\\
\end{eqnarray}

\noindent Sheldon \cite{SHELDON64} obtained

\begin{equation}
\eta_{\ell-1}-\eta_{\ell+1}=-\frac{\pi}{2}\int_0^{\infty}\left[J_{\ell-1/2}(kr)\right]^2V(r)rdr+\frac{\pi}{2}\int_0^{\infty}\left[J_{\ell+1/2}(kr)\right]^2V(r)rdr,
\end{equation}

\noindent which corresponds the the phase-shift difference in the Born approximation. Therefore, for instance 

\begin{eqnarray}
& &\int_0^{\infty}x^{\lambda-1}e^{-\alpha x}\left[J_{\ell+1/2}(kx)\right]^2dx\nonumber\\
& &\;\;\;\;\;\;\;\;\;\;\;\;\;\;=\frac{k^{2\ell+1}}{\sqrt{\pi}\alpha^{2\ell+1+\lambda}}\frac{\Gamma(\ell+1)\Gamma(2\ell+1+\lambda)}{\Gamma(\ell+\frac{3}{2})\Gamma(2\ell+2)}\nonumber\\
& &\;\;\;\;\;\;\;\;\;\;\;\;\;\;\times~_3F_2\left[\begin{array}{c}
\ell+1,\ell+\frac{\lambda+1}{2},\ell+1+\frac{\lambda}{2}\\
\ell+\frac{3}{2}, 2(\ell+1)
\end{array};-\frac{4k^2}{\alpha^2}\right],
\end{eqnarray}

\noindent where $\Gamma$ is the usual Gamma function and $_pF_q$ is a generalized hypergeometric function defined in Eq. (\ref{def_fpq}).

\end{document}